\documentclass[10pt,
               aps,
               pre,
               showpacs,
               showkeys,
               twocolumn,
               ]{revtex4}
\usepackage{isolatin1}
\usepackage{graphicx}
\usepackage{amsmath}
\usepackage{amssymb}

\begin{document}

\title{Hypothesis of strong chaos and anomalous diffusion in coupled
  symplectic maps}
\author{Eduardo G. Altmann}
\affiliation{Max Planck Institute for the Physics of Complex Systems, N\"othnitzer Strasse 38, 01187 Dresden, Germany.} 
\author{Holger Kantz}
\affiliation{Max Planck Institute for the Physics of Complex Systems, N\"othnitzer Strasse 38, 01187 Dresden, Germany.}

\date{\today}

%\date{\today}
%----------------------\input{Abstract.tex}---------------------------------------

\begin{abstract}
We investigate the high-dimensional Hamiltonian chaotic dynamics in~$N$ coupled
area-preserving maps. We show 
the existence of an enhanced trapping regime caused by trajectories performing a random walk {\em inside} the area
corresponding to 
regular islands of the uncoupled maps. As a consequence, we observe long 
intermediate regimes of power-law decay of the recurrence time statistics (with
exponent~$\gamma=0.5$) and of ballistic motion.  The asymptotic decay of correlations   and anomalous
diffusion depend on the stickiness of the $N$-dimensional invariant tori.  Detailed numerical
simulations show weaker stickiness for increasing~$N$ suggesting
that such paradigmatic class of Hamiltonian systems asymptotically fulfill the demands of the usual
hypotheses of strong chaos.
\end{abstract}


\maketitle

\section{Introduction}\label{sec.introduction}
Hypotheses of strong chaos are keystone requisites for both basic theories  (statistical mechanics~\cite{dorfman,gallavotti})
and computation methods (transport properties~\cite{dorfman}, elimination of chaotic 
variables~\cite{riegert.05}).
In all cases, the hypotheses of strong chaos consist essentially of
two main assumptions about the (high-dimensional Hamiltonian) dynamics: (i) ergodicity, i.e., the existence of a single chaotic
component and negligible measure of the regions of quasi-periodic motion; and
(ii) strong mixing (exponential decay of
correlations). Generic low-dimensional Hamiltonian systems violate both
hypotheses since the chaotic trajectories stick to the border of islands of regular 
motion introducing
long-term correlations in the dynamics~\cite{meiss.ott.prl,zas.pr,chirikov.prl}.

For many higher-dimensional systems it is generally expected that  hypothesis (i)
is effectively valid since the measure of the quasi-periodic trajectories
typically decreases (exponentially) with the number~$N$ of degrees of freedom
(see, e.g., Sec. 6.5 of ref.~\cite{lichtenberg}).
However, the vanishing measure of regular 
regions [hypothesis (i)] does not guarantee the fast decay of correlations
[hypothesis (ii)]. The breakdown of hypothesis (ii)
occurs due to the trapping (stickiness) of chaotic trajectories around non-hyperbolic
structures in the phase-space, and even zero measure sets (e.g., the bouncing  ball orbits in the
Bunimovich stadium billiard) can be responsible for the anomalous decay of 
correlations~\cite{altmann.sharp,armstead}.   Considerable progress has been achieved for the
problem of trapping in area-preserving maps
$(N=1)$~\cite{meiss.ott.prl,chirikov.prl,zas.pr}, while only few
numerical results are known for~$N=2$--$5$~\cite{ding.90,astronomy}. On the
other hand, higher-dimensional systems are usually considered for the
calculation of different properties, such as relaxation
phenomena~\cite{latora.99} and diffusion~\cite{kaneko.89,boffetta}.

% or calculating the spectrum of Lyapunov exponents~\cite{falcioni.91}. 

In this Letter we perform a detailed study of the trapping properties of 
Hamiltonian systems with an increasing number of degrees of freedom by
coupling~$N$ area-preserving maps~\cite{ding.90,kaneko.89}. This system
  typically has invariant tori, but hypothesis
  (i) is satisfied for large~$N$ in the sense mentioned above 
(our results do  not apply when regular behavior prevails for~$N\rightarrow\infty$~\cite{referee}).
For weak coupling we observe an intermediate regime of enhanced trapping which we show to exist also
for area-preserving maps perturbed by noise. Asymptotically, our numerical
results show that the trapping decreases with~$N$ 
indicating the effective validity of hypothesis~(ii) for high-dimensional
Hamiltonian systems composed of coupled low-dimensional ones. Additionally,
we show how the different trapping regimes impact on the anomalous diffusion
of the perturbed standard map and lead to a non-trivial dependence of the
asymptotic diffusion coefficient on the perturbation strength, clarifying previous
conflicting results~\cite{kaneko.89,ishizaki,floriani,boffetta}.

\section{Coupled symplectic maps}\label{sec.model}
We construct a time-discrete 
$2N$-dimensional Hamiltonian system by the composition~$\boldsymbol{T} \circ \boldsymbol{M}$ 
of the independent one-step iteration of $N$
symplectic 2-dimensional maps $\boldsymbol{M}=(M_1,\ldots,M_N)$ and a
symplectic coupling $\boldsymbol{T}=(T_1,\ldots,T_N)$.  
As a representative example of $2$-d maps we
choose for our numerical investigation the standard
map~\cite{lichtenberg}: 
\begin{equation}\label{eq.standard}
M_i \left( \begin{array}{ll}
p_i \\ q_i
\end{array}
\right)=\left(
\begin{array}{ll}
p_i +K_i \sin(2\pi q_i) \; &\text{mod 1}\\
q_i+p_i +K_i \sin(2\pi q_i)  &\text{mod 1}\\
\end{array}
\right),
\end{equation}
and a coupling potential between the maps~$i,j$ given by~$V_{i,j}=\xi_{i,j}
\cos[2\pi(q_j-q_i)]$\cite{f1}.
The action of the coupling on the $i$-th map is hence given by
\begin{equation}\label{eq.t1}
T_i  \left( \begin{array}{ll}
p_i\\ q_i
\end{array} \right)= \left( \begin{array}{ll}
p_i+\sum_{j=1}^{N} \xi_{i,j} \sin[2\pi(q_i-q_j)] \\
q_i\\
\end{array}
\right),
\end{equation}
which corresponds to a perturbation~$\Delta p_i$.
The full coupling~$\boldsymbol{T}$ is symplectic provided
$\xi_{i,j}=\xi_{j,i}$.
For simplicity we use all-to-all coupling
with~$\xi_{i,j}=\frac{\xi}{\sqrt{N-1}}$, in which case a numerically
convenient mean-field representation can be written~\cite{latora.99}
\begin{equation}\label{eq.hsm}
\Delta p_i=\frac{\xi}{\sqrt{N-1}} \sum_{j=1}^{N}
\sin[2\pi(q_i-q_j)]= \xi |\boldsymbol{m}| \sin(2\pi q_i-\phi),
\end{equation}
where
$\boldsymbol{m}=(m_x,m_y)=\frac{1}{\sqrt{N-1}}\sum_j[\cos(q_j),\sin(q_j)]$
and~$\tan(\phi)=m_y/m_x$. When the isolated systems are chaotic and weakly 
coupled we can assume that the positions~$q_j$ are uncorrelated and approximate
each term of the sum 
in~Eq.~(\ref{eq.hsm}) by a random variable~$y\in]-1,1[$ distributed according
    to the density $P(y)=1/(2\pi\sqrt{1-y^2})$, which has
    variance~$\sigma_y=\sqrt{2}/2$. Additionally, for large~$N$ Eq.~(\ref{eq.hsm}) tends
    to a normal distribution with $\sigma=\xi \sigma_y$, i.e., a finite
    perturbation strength. This is the main motivation for the choice of the
    rescale factor~$1/\sqrt{N-1}$ what is a major difference from other
    coupled-oscillators models, where a factor $1/N$ is used instead~\cite{referee,latora.99}. 

\section{Trapping in the model}\label{sec.trapping}

%\subsection{Standard map perturbed by noise}\label{ssec.noise}

\begin{figure}[!ht]
\centerline{
\includegraphics[width=\columnwidth]{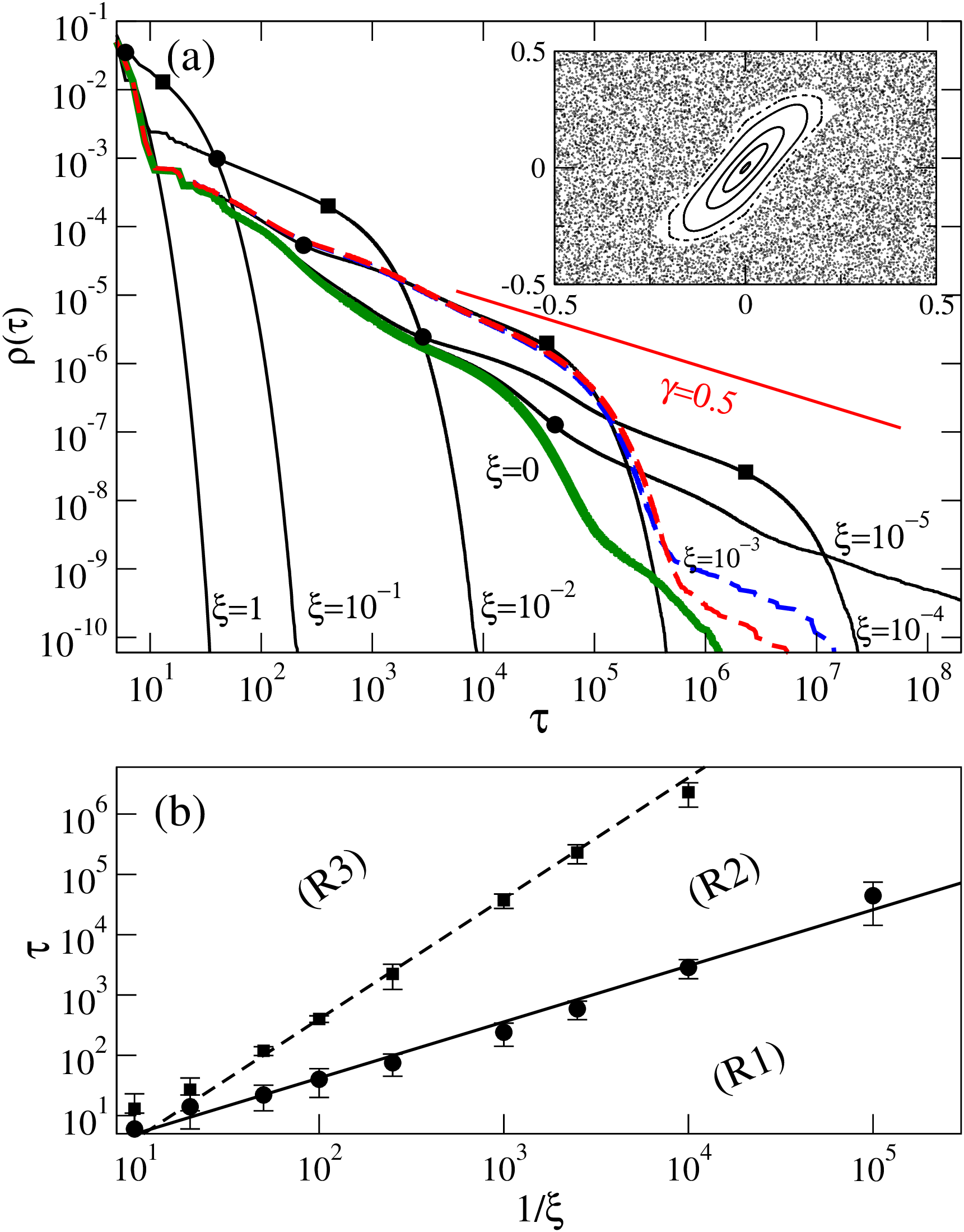}}
\caption{(Color online) (a) RTS for the standard map with $K=0.52$. The
   thick solid line corresponds to the unperturbed case ($\xi=0$), whose phase
   space is shown in the inset. The thin solid lines correspond to
   perturbations due to white noise with~$\xi=10^{0},...,10^{-5}$ (from left
   to right). In each curve $\tau_{1,2}$ ($\bullet$) and~$\tau_{2,3}$
   ({\tiny $\blacksquare$}) are indicated. The dashed lines
   correspond to  the cases of two (upper) and three (lower) coupled standard-maps
   with~$\xi=10^{-3}, K_1=0.5214,$ and $K_2=0.5108$~\cite{f2}. (b) Dependence
   of~$\tau_{1,2}$ 
   and~$\tau_{2,3}$  on~$\xi$:  $\tau_{1,2}=0.58
   \xi^{-0.93}$ (lower line) in agreement with
   Eq.~(\ref{eq.tau12}); and $\tau_{2,3}=0.04 \xi^{-2}$ (upper line) in
   agreement with Eq.~(\ref{eq.tau23}).}
\label{fig.1}
\end{figure}

Motivated by the previous considerations we study initially the trapping
properties of one standard
map ($N=1$) perturbed by white noise, i.e, we
replace Eq.~(\ref{eq.hsm}) by~$\Delta p_1=\xi \delta$, where~$\delta$ is a Gaussian
distributed random number with zero mean and variance~$\sigma=\sigma_y$.
We fix~$K_1=0.52$ in~(\ref{eq.standard}), where a single large regular island is visible in
the phase space (inset of fig.~\ref{fig.1}), what facilitates the
interpretation of our results. The trapping is measured by regimes of power-law decay of the recurrence time statistics
(RTS)~$\rho(\tau)\sim\tau^{-\gamma}$, defined as the 
probability of a trajectory to return at a time~$T>\tau$ to a pre-defined
region. During the long recurrence times the trajectory performs
almost a quasi-periodic motion leading to a power-law decay of correlations
with exponent~$\gamma_c=\gamma-1$~\cite{chirikov.prl}. 
 We have iterated a single trajectory~$10^{12}$ times and recorded the
 times~$T$ 
between successive recurrences to a large region away from the island~\cite{f1.5}.
 The RTS for different noise
intensities~$\xi$ is shown in fig.~\ref{fig.1}. Three different regimes can 
be identified for small~$\xi$: %(e.g., in the case~$\xi=10^{-3}$):

{\bf (R1)} For short times~$(\tau<\tau_{1,2})$ the RTS follows
the unperturbed one ($\xi=0$), i.e., it shows an 
exponential followed by a power-law decay (with exponent~$\gamma_{R1}$).   

{\bf (R2)} For intermediate times $(\tau_{1,2}<\tau<\tau_{2,3})$ the RTS
shows an enhanced trapping due to trajectories that entered the island
through the action of the noise. Once inside the island the trajectories
revolve around the central elliptic periodic orbit and perform a random walk
in the perpendicular direction. The power-law exponent tends to the
value  of a random
walker~$\gamma_{R2} \approx \gamma_{RW}=0.5$. As in typical RTS of Hamiltonian systems, we observe 
additional oscillations.

{\bf (R3)} For long times $(\tau>\tau_{2,3})$  the RTS decays
exponentially.

We obtain now the dependence of~$\tau_{1,2}$ and~$\tau_{2,3}$ on~$\xi$.
The starting time of (R2), $\tau_{1,2}$, occurs roughly when the
displacement due to the noise is
of the same size of the distance to the regular island. Following the
arguments of ref.~\cite{floriani} one obtains
\begin{equation}\label{eq.tau12}
\tau_{1,2} \sim \xi^{-\beta}.
\end{equation}
Numerical simulations in different maps show that $\beta \lessapprox 1$. In
ref.~\cite{floriani} the noise intensity was compared to the size of the 
chaotic layer between two cantori close to the island.  Using the Markov-tree model for stickiness introduced in
ref.~\cite{meiss.ott.prl} it was obtained that $\beta=1/(2\gamma_{R1}-1)$. Since usually
$1<\gamma_{R1}<2$ one obtains $1/3 < \beta< 1$. Differently from 
ref.~\cite{floriani}, we do not observe an exponential decay right after this
time. As argued above, the exponential regime (R3) appears at a later
time~$\tau_{2,3}>\tau_{1,2}$ due to the finiteness of the random walk domain. 
Considering the
measure~$\mu_I$ of the (largest) island inside which a random walker (with step size proportional
to~$\xi$) performs a diffusive motion, the dependence of~$\tau_{2,3}$ on~$\xi$ can be estimated as
\begin{equation}\label{eq.tau23}
%sqrt{\mu_I} \sim \xi \sqrt{\tau_{2,3}} \Rightarrow 
\tau_{2,3} \sim \mu_I  \xi^{-2}.
\end{equation}
In fig.~\ref{fig.1}b we verify the agreement of 
Eqs.~(\ref{eq.tau12}) and (\ref{eq.tau23}) with the values estimated
numerically. The numerical values of $\tau_{1,2}$ and~$\tau_{2,3}$  were
  estimated as the intersecting point of two fitted power-laws (or
  exponential), which extend over more than two decades for small~$\xi$ and can be extrapolated for higher~$\xi$. 
The noise has two qualitatively different effects: while (R3) represents the typical cut-off of the power-law
distribution~\cite{floriani}, during the novel regime (R2) the noise 
acts constructively (increasing the regularity of the dynamics) by allowing 
trajectories to penetrate regular islands.

%\subsection{Higher dimensional Hamiltonian system}\label{ssec.higher}

We investigate now the fully deterministic system given 
by the composition of Eqs.~(\ref{eq.standard}) and~(\ref{eq.t1}). The
existence of $N$-dimensional invariant tori is confirmed for small
coupling by the application of the Kolmogorov-Arnold-Moser theorem to the tori built as a
direct product of the~$1$-dimensional tori of the $N$ uncoupled maps 
(the quasi-periodic orbits inside the island).
All our numerical simulations for $\xi\leq0.1$ are
consistent with the assumption that these are the only existent tori, in agreement with a more general picture of Hamiltonian systems: exponential
decay of the measure of the tori with~$N$ ($\mu_{tori}={\mu_I}^N$ 
for~$\xi\rightarrow0$) and nonexistence of tori of dimension smaller than~$N$
-- ``Froeschl\'e conjecture''~\cite{lichtenberg}. Furthermore, due to Arnold
diffusion, it is consistent to assume that a single chaotic ergodic
component exists. 
This means that the motion in the~$2N$-dimensional phase space
belongs to 
{\em the} chaotic component if the variables of at least one map
belong to its chaotic component when uncoupled.

\begin{figure}[!ht]
\centerline{
\includegraphics[width=\columnwidth]{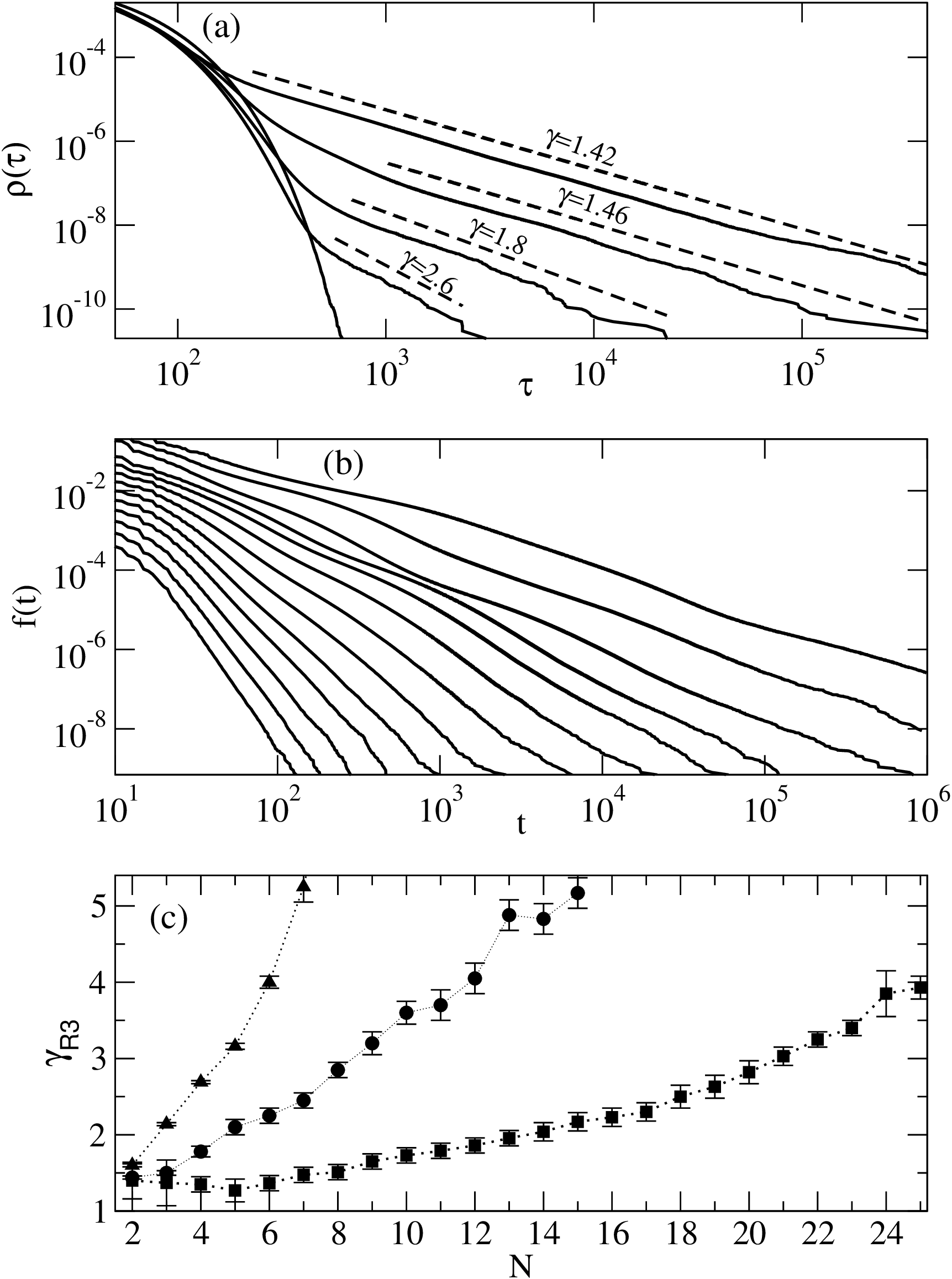}}
\caption{(a) RTS for $\xi=0.05$ and $N=2,3,4,5,$ and
  noise (from top to bottom). (b) Survival probability~$f(t)$ of $10^{10}$
  trajectories near the~$N$-dimensional tori for~$\xi=0.05$ and $N=2,...,15$
  (from top to bottom). (c) Exponent~$\gamma_{R3}$ fitted to the power-law
  regime of~$f(t)$ for~$\xi=0.1,0.05,0.03$ (from top to bottom). In all cases~$K_1=0.52$ and~$K_i\in[0.51,0.53],
  i\in\{2,...,N\}$ were used in map~(\ref{eq.standard}).}
\label{fig.2}
\end{figure}

 The numerical RTS obtained for~$N>1$ and for the noisy model are almost
 indistinguishable for recurrence times belonging to the regimes~(R1)
 and~(R2) described above. Two representative examples -- for~$\xi=10^{-3}$ and~$N=2,3$ -- are depicted as dashed lines
in fig.~\ref{fig.1}\cite{f2}.  Remarkable differences are observed in
 regime~(R3). The detailed analysis shown in fig.\ref{fig.2}a for~$\xi=0.05$
 indicates that: 

{\bf (R3')} For large times $(\tau>\tau_{2,3})$  the RTS shows an
exponential followed by a power-law decay (with exponent~$\gamma_{R3}$) due to the stickiness to~$N$-dimensional tori.

We investigate next the dependence on~$N$ of the asymptotic power-law
exponent~$\gamma_{R3}$, which corresponds to a pure high-dimensional
effect. The term asymptotic here has to be taken with caution since already for~$N=1$
the convergence of the RTS to a well defined power-law is very slow in $\tau$~\cite{chirikov.prl}. 
However, for our purposes it is enough to perform a comparative analysis for
different~$N$ and similar recurrence times and
initial conditions.
It is necessary to distinguish between two effects of~$N$ on the RTS 
seen in fig.~\ref{fig.2}a: the later onset of the power-law
regime for increasing~$N$, related to the smaller measure of the tori --
hypothesis~(i); and the different values
of~$\gamma_{R3}$ (slopes of the tails), related to the stickiness to higher-dimensional
tori -- hypothesis (ii) investigated here. Figure~\ref{fig.2}a already
suggests that $\gamma_{R3}$
increases with~$N$. In ref.~\cite{ding.90} similar results were reported for a
different system and~$N=2$ and~$3$. For improved statistics in the tails we
study the survival probability~$f(t)$ inside a region containing the
$N$-dimensional tori of $10^{10}$ trajectories started {\em close} to them~\cite{f3}.
$f(t)$ is shown in fig.~\ref{fig.2}b where it is evident
that the power-law exponent increases with~$N$. The estimated exponents
shown in fig.~\ref{fig.2}c for different moderate values of the coupling~$\xi$ suggest a linear dependence~$\gamma_{R3} \propto
N$. This result is consistent with hypothesis~$(ii)$ for high-dimensional
systems since the sharp tails indicate fast decay of
correlations.

\section{Anomalous diffusion}\label{sec.diffusion}

One of the most important effects induced by the stickiness is the anomalous
 diffusion of chaotic trajectories. %~\cite{swinney}. 
 We consider next the influence of the trapping regimes~(R1-3) discussed above on the
widely studied case of diffusion in the momentum~$p$ of the standard
map~\cite{lichtenberg,zumofen,zas.pr}. The diffusion is measured 
 considering the map~(\ref{eq.standard}) opened in~$p$ and calculating the temporal evolution of the
 dispersion of an ensemble of trajectories: $\langle \Delta p^2(t) \rangle =
 D t^\nu.$ 
Diffusion is normal if~$\nu=1$. 
It is well-known that anomalous superdiffusion~$1<\nu<2$ is obtained for the
parameters $K$ of the standard map where the so-called accelerator modes
(ballistic islands of the open map) exist~\cite{lichtenberg,zumofen,zas.pr}.
The diffusion exponent~$\nu$ is directly linked to the
exponent~$\gamma$ (of the RTS due to the trapping around accelerator modes) by~\cite{zumofen}
\begin{equation}\label{eq.rectransp}
\nu = \left\{ \begin{array}{lll}
    2    & \text{ if } \;  \gamma < 1\;,\\
3-\gamma & \text{ if } \; 1 \leq \gamma \leq 2\;, \\
1        & \text{ if } \; \gamma\;>2. \\
\end{array}
\right.
\end{equation}

Previous publications report numerical results that emphasize different
effects of the noise on the diffusion: while
in refs.~\cite{floriani,boffetta} the onset of normal diffusion was obtained,
in ref.~\cite{ishizaki} the possible enhancement of anomalous diffusion
was reported. Applying the results described above we 
show that actually both effects exist for different time scales.

\begin{figure}[!ht]
\centerline{
\includegraphics[width=\columnwidth]{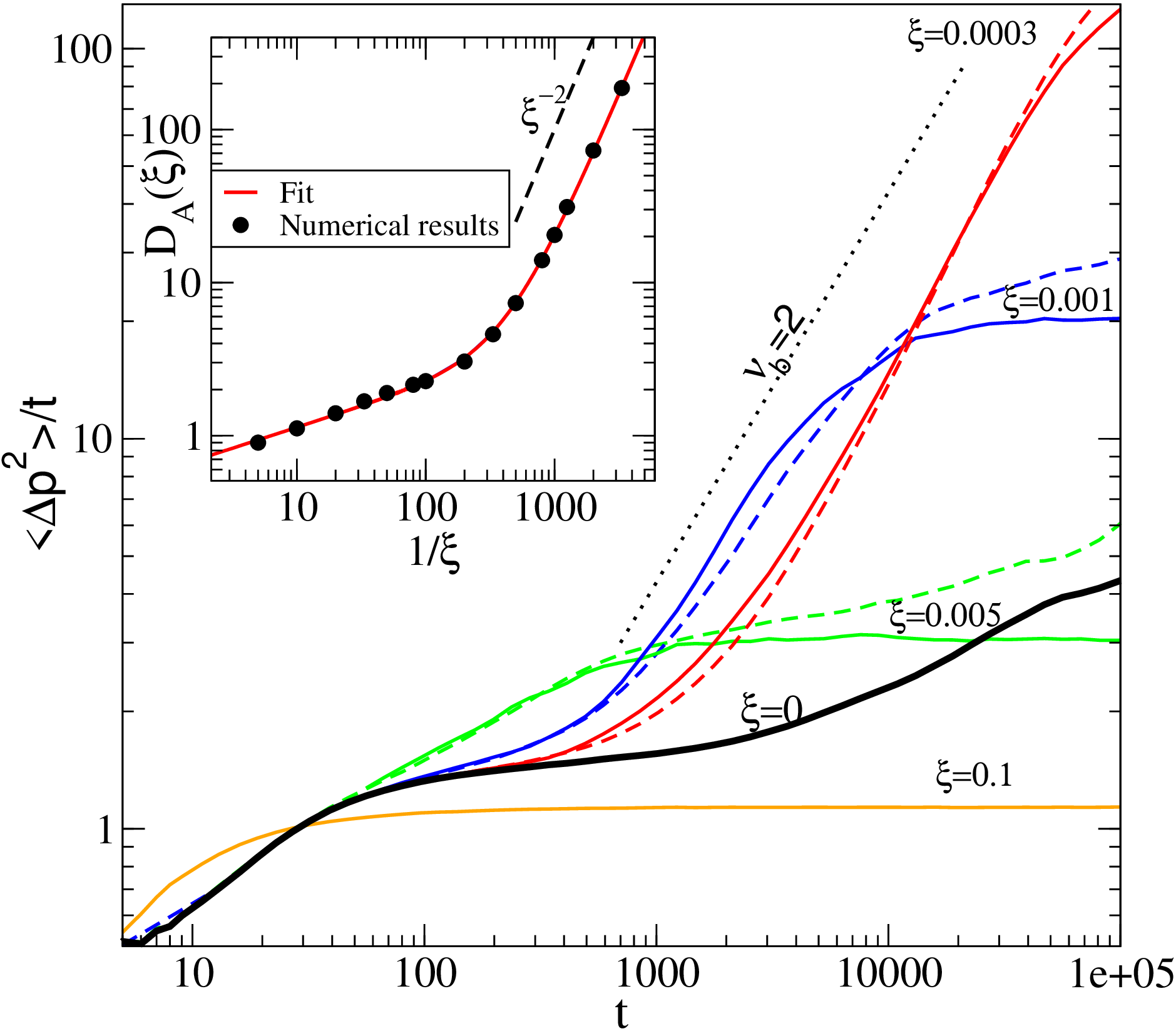}}
\caption{ (Color online) Diffusion in a standard map with~$K_1=1.07$
  (accelerator mode) coupled to noise
  (thin solid lines) and to another standard map with~$K_2=0.52$ (dashed
  lines). The uncoupled case is depicted as a thick solid line. From top to
  bottom~$\xi=0.0003,0.001,0.005,0.1$. $10^6$ trajectories were
  used with initial conditions away from islands. When the curves become
  constant (for $t\rightarrow \infty$) diffusion is normal. Inset: asymptotic diffusion coefficient for different noise
  intensities. The solid line corresponds to Eq.~(\ref{eq.D12}) with
  $a=0.604, b=1.65\;\cdot 10^{-5}$ and $-\beta(1-\nu)=0.273$.}  
\label{fig.3}
\end{figure}

The three trapping regimes (R1-3) induce different diffusion regimes,
as shown in fig.~\ref{fig.3} and described below. During (R1) the anomalous  
diffusion is similar to the unperturbed
case. In (R2) the trapping exponent tends to~$\gamma_{R2}=0.5$ and ballistic
motion~$\nu=2$ is predicted 
according to Eq.~(\ref{eq.rectransp}). We have verified numerically that the
beginning and end of this regime occur at 
times proportional to, but greater than, those of (R2) (we denote these times
by $\tau_{1,2}^\dagger$ and~$\tau_{2,3}^\dagger$, respectively). 
Asymptotically the exponential RTS in (R3) implies normal
diffusion~\cite{f4}. The asymptotic diffusion coefficient~$D_A=\lim_{t\rightarrow\infty} \langle 
\Delta p^2(t)\rangle/t$ can be determined by the intermediate anomalous regimes and
shows a nontrivial dependence on the
noise intensity~$\xi$: for weak noise,
Eqs.~(\ref{eq.tau12}) and~(\ref{eq.tau23}) indicate that the dominant contribution
comes from the ballistic regime associated to~(R2) and thus $D_A \approx [D_{R2}
  (\tau^\dagger_{2,3}-\tau^\dagger_{1,2})] \sim \xi^{-2}$. For stronger noise
  the major contribution is given by a regime of
  superdiffusion corresponding approximately to (R1),
  that can be estimated as~$D_A \approx (D_{R1}
  \tau^\dagger_{1,2})^{\nu-1}\sim \xi^{-\beta(\nu-1)}$, where $\nu$ is the  
  unperturbed anomalous diffusion exponent (for small times).
%which is related
 % to~$\beta$ and $\gamma$ through Eqs.~(\ref{eq.tau12})
 % and~(\ref{eq.rectransp}). 
Considering the composition of this two effects
  and joining the multiplicative terms in two fitting parameters~$a,b$ we obtain 
\begin{equation}\label{eq.D12}
D_A(\xi)= a \; \xi^{-\beta(\nu-1)}+b\; \xi^{-2}.
\end{equation}
In the inset of fig.~\ref{fig.3} we show the remarkable
agreement of the numerically obtained diffusion coefficient and
expression~(\ref{eq.D12}).
Specially interesting is the transition for small~$\xi$ to an
asymptotic~$\xi^{-2}$ dependence of~$D_A$, which can be seen as a direct consequence of
the nontrivial trapping regime~(R2), is  
absent in the case of 1-d maps (e.g., Pomeau-Manneville
maps)~\cite{floriani,bettin}, and was not previously
reported in refs.~\cite{boffetta,floriani,ishizaki}. The asymptotic~$\xi^{-2}$
scale was predicted through different arguments in ref.~\cite{karney}, and can be traced back to the works of Taylor~\cite{taylor}.

The diffusion theory described above is valid for noise perturbation and for
deterministic high-dimensional systems whenever~$\gamma_{R3}>2$ [see
  Eq.~(\ref{eq.rectransp})]. When~$\gamma_{R3}<2$ one observes
additionally an
asymptotic regime of anomalous diffusion, as shown in
fig.~(\ref{fig.3}) for~$N=2$ coupled standard maps and different coupling strengths~$\xi$. We see that
the question of whether the asymptotic diffusion is anomalous or normal 
depends crucially on how the chaotic trajectories stick to~$N$-dimensional tori
(exponent~$\gamma_{R3}$). Our results indicate that in
general~$\gamma_{R3}$ 
increases with~$N$ (fig.~\ref{fig.2}) and the diffusion is
normal, in agreement with refs.~\cite{boffetta,kaneko.89}. However, our
explanation for this behavior is {\em not} the absence of hierarchical tori or the Arnold
diffusion, as argued in ref.~\cite{kaneko.89} for a similar system, but 
simply that~$\gamma_{R3}>2$ for sufficiently large~$N$.

\section{Conclusion}\label{conclusion}

We have shown that correlations decay faster as the 
dimensionality~$N$ of a paradigmatic class of Hamiltonian systems increases. More precisely, our numerical
simulations show that the asymptotic power-law exponent of the RTS increases 
with~$N$, providing evidence that, for large enough~$N$ and times, coupled
symplectic maps can be considered for all practical purposes as ergodic and
strongly chaotic. This suggests a novel explanation for the onset of strong
chaos in high-dimensional Hamiltonian systems whose generality has to be
determined through the specific study of other classes of systems. An
unexpected behavior is that for small coupling strength or noise intensity a
long intermediate enhanced-trapping 
regime with~$\gamma_{R2}\approx0.5$ exists due to the trapping inside
remnants of lower-dimensional regular regions. This implies a
regime of enhanced anomalous diffusion and a non-trivial dependence of the
asymptotic diffusion coefficient on the perturbation strength, given by
Eq.~(\ref{eq.D12}). Applications of these results include
noise-perturbed low-dimensional system as well as fully deterministic examples
such as galaxy dynamics~\cite{astronomy} and active
  transport~\cite{boffetta}. Further investigations are needed to determine
whether our results provide a valid description also of systems like
Hamiltonian mean field models, where similar transient regimes were
observed~\cite{latora.99}.  

\acknowledgments
E.G.A. thanks A.E.Motter, N. Baba, and G. Cristadoro for helpful discussions and CAPES (Brazil) for financial support.


\begin{thebibliography}{19}
\bibitem{f1}
{We have verified our main results using different area-preserving maps and coupling potentials.}
\bibitem{f1.5}
{The size and position of the recurrence region determine the
  exponent of the short-time exponential decay of the RTS and has no influence in
  the power-law regimes. We have used $|q_1|>0.25$ as
  recurrence region.}
\bibitem{f2}
{In order to avoid artificial symmetries we choose~$K_i\in[0.51,0.53], i\in\{2,...,N\}$
and fix~$K_1=0.52$. To
facilitate the comparison with the RTS of the noisy system we use the
projection into $(q_1,p_1)$ and the same recurrence region. Similar
results are obtained in the full phase space.}
\bibitem{f3}
{Inside the
island~$q_i\in[0.45,0.55],p_i \in [0.4,0.6]$ for~$i\in\{2,...,N\}$ and 
close to it~$q_1=0.65,p_1 \in [0.835,0.845]$ in the remaining map.}
\bibitem{f4}
{While for the intermediate regimes it is essential that initial conditions are
  taken outside the accelerator modes, due to the ergodicity induced by the noise the asymptotic diffusion coefficient is
  independent of the choice of initial conditions. }
\end{thebibliography}
\end{document}